\documentclass[12pt,epsf]{article}
\usepackage{amsmath}
\usepackage{amsthm, amssymb}
\usepackage{graphicx}
\usepackage{setspace}
\usepackage{subfigure}
\usepackage{cite}

\theoremstyle{remark}

\newcommand{\be}{\begin{equation}}
\newcommand{\ee}{\end{equation}}
\newcommand{\bea}{\begin{eqnarray}}
\newcommand{\eea}{\end{eqnarray}}
\newcommand{\bear}{\begin{eqnarray}}
\newcommand{\eear}{\end{eqnarray}}
\newcommand{\beas}{\begin{eqnarray*}}
\newcommand{\p}{\partial}
\newcommand{\eeas}{\end{eqnarray*}}
\newcommand{\ba}{\begin{array}}
\newcommand{\ea}{\end{array}}

\newcommand{\ra}

\newcommand{\pd}[2][1]{\ifnum#1=1 \frac{\partial}{\partial {#2}} \else
  \frac{\partial^#1}{\partial {#2}^{#1}}\fi}
\newcommand{\dpd}[2][1]{\ifnum#1=1 \dfrac{\partial}{\partial {#2}} \else
  \frac{\partial^#1}{\partial {#2}^{#1}}\fi}
\newcommand{\td}[2][1]{\ifnum#1=1 \frac{d}{d{#2}} \else
  \frac{d^#1}{d{#2}^{#1}}\fi}

\newcommand{\nbox}{{\,\lower0.9pt\vbox{\hrule \hbox{\vrule height 0.2 cm \hskip 0.19 cm \vrule height 0.2 cm}\hrule}\,}}

\def\href#1#2{#2}

\usepackage{color}

\usepackage{xcolor}
\usepackage[citebordercolor=green, linkbordercolor={ 0 0 1}, linktocpage=true]{hyperref}

\begin{document}
\begin{titlepage}
\begin{NoHyper}
\hfill
\vbox{
    \halign{#\hfil         \cr
           } 
      }  
\vspace*{20mm}
\begin{center}
{\Large \bf Towards a Reconstruction of General Bulk Metrics}

\vspace*{15mm}
\vspace*{1mm}
Netta Engelhardt and Gary T. Horowitz
\vspace*{1cm}
\let\thefootnote\relax\footnote{engeln@physics.ucsb.edu, gary@physics.ucsb.edu}

{Department of Physics, University of California\\
Santa Barbara, CA 93106, USA}

\vspace*{1cm}
\end{center}
\begin{abstract}
We prove that the  metric of a general holographic spacetime can be reconstructed  (up to an overall conformal factor) from distinguished spatial slices -- ``light-cone cuts'' -- of the conformal boundary. Our prescription is covariant  and applies to bulk points in causal contact with the boundary. Furthermore, we  describe a procedure for determining the light-cone cuts corresponding to bulk points in the causal wedge of the boundary  in terms of the divergences of correlators in the dual field theory. Possible extensions  for determining the conformal factor and including the cuts of points outside of the causal wedge are discussed. We also comment on implications for subregion/subregion duality.

\end{abstract}
\end{NoHyper}

\end{titlepage}
\tableofcontents
\vskip 1cm
\begin{spacing}{1.2}
\section{Introduction}\label{sec:intro}

One of the most intriguing aspects of  gauge/gravity duality~\cite{Mal97,Wit98a, GubKle98} is its implication that spacetime geometry is emergent. The metric is not a fundamental variable of quantum string theory with asymptotically Anti-de Sitter (AdS) boundary conditions: it is rather an object which emerges in the appropriate limits. The quantum structure from which spacetime emerges, however, remains mysterious. There is an active research program devoted to reconstruction of the bulk metric from the dual field theory.

Much of this program has focused on recovering the bulk geometry from various measures of quantum entanglement in the dual field theory (starting with~\cite{Van09, Van10}). This approach is particularly appealing since entanglement entropy is dual to the area of bulk extremal surfaces
~\cite{RyuTak06, HubRan07}:  it may be possible to reconstruct the metric by comparing  entanglement entropy for different regions. To our knowledge, the most developed approaches along these lines are ``hole-ography'' and the (related) construction of kinematic space~\cite{ BalChoCze, MyeHea14, CzeLam, CzeLamMcC15a, BalCzeCho, CzeDonSul, MyeRaoSug}. However, this method of bulk reconstruction suffers from some drawbacks: it is at this time understood in only a limited class of cases, and in particular it is not formulated for generic holographic spacetimes in more than $2+1$ dimensions\footnote{Generalizations to higher dimensions are limited to highly symmetric setups~\cite{MyeRaoSug}.}, and  it is subject to a set of no-go theorems and constraints discussed in~\cite{Hub12, EngWal13, EngFis15}. Other approaches to reconstruction, e.g.~\cite{deHSolSke00, HamKab06, Kab11, ChrSke16} 
often assume the bulk equations of motion. 

 We provide an alternative approach for reconstructing the bulk metric from field theory data.    The reconstruction we propose is based on a  new way of identifying bulk points in terms of distinguished boundary spatial slices, the ``light-cone cuts''. We will give a complete prescription for recovering the  bulk conformal metric, i.e. the metric up to an overall conformal rescaling, just from the location of the light-cone cuts. (We believe it should also be possible to  obtain the conformal factor, but this is still under investigation.)  We will then show that the light-cone cuts themselves may be found  from the divergence structure of boundary $n$-point functions, using the work of~\cite{MalSimZhi}. 
This approach is completely well defined for (most) bulk points in the causal wedge of the entire asymptotic boundary, i.e. points which have both past and future causal contact with the asymptotic boundary. As we will discuss, the former part of the reconstruction is also valid outside of this region, including some points inside  a black hole event horizon; it is not yet clear how to extend the latter part.
 We emphasize that our approach  is covariant and well-defined for any holographic spacetime of any dimension. We make no assumptions about the matter content; in particular, we do not assume the null energy condition.

\begin{figure}[t]
\centering
\includegraphics[width=8cm]{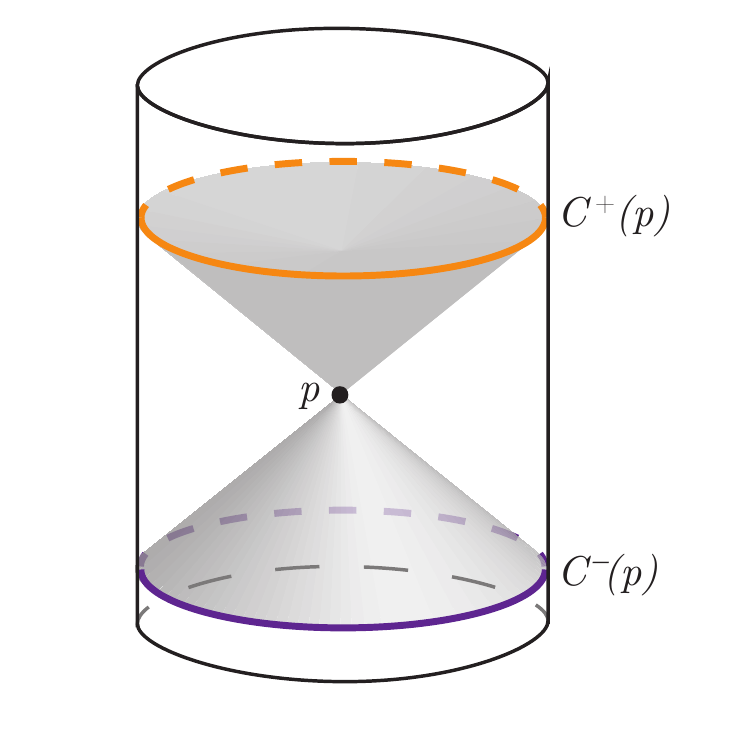}
\caption{The intersection of the lightcone (up to caustics) of a bulk point $p$ with the asymptotic boundary defines the past and future cuts of $p$, $C^{\pm}(p)$. The cuts are complete spatial slices of the asymptotic boundary.}
\label{fig:pfcutIntro}
\end{figure} 

The starting point of our procedure is the construction of a unique spatial slice of the boundary geometry from any bulk point, provided that the point is within causal contact of the boundary.  This slice is the intersection of the (past or future) light cone of the bulk point with the asymptotic boundary (up to caustics), as illustrated in Fig.~\ref{fig:pfcutIntro}. Clearly, not every boundary spatial slice corresponds to the light cone of a bulk point. We call the special slices which do correspond to bulk points ``light-cone cuts'', or ``cuts'' for short. Our approach is similar in spirit to the program initiated by Newman in the 1970's \cite{New76, HanNewPen} for asymptotically flat spacetimes. In particular, it was shown in \cite{KozNew83} that the conformal metric of an asymptotically flat spacetime can be recovered from similar light-cone cuts at null infinity.  A crucial new ingredient in our approach is the use of the dual field theory to determine the cuts.

We show that a  light-cone cut corresponds to a unique bulk point, and give a prescription for reconstructing the conformal metric from the set of light-cone cuts. In this way, we show that the space of cuts acts as an auxiliary spacetime, filling a similar role in this causal reconstruction as de Sitter space does in the geodesic reconstruction of~\cite{CzeLamMcC15a}. Our approach to reconstruction is local: we determine the conformal metric pointwise. We also present partial results for determining the causal separation for points at finite separation directly from the behavior of their cuts.
The causal relations between certain points can be determined  from the type of intersection of their cuts.

\begin{figure}[t]
\centering
\includegraphics[width=8cm]{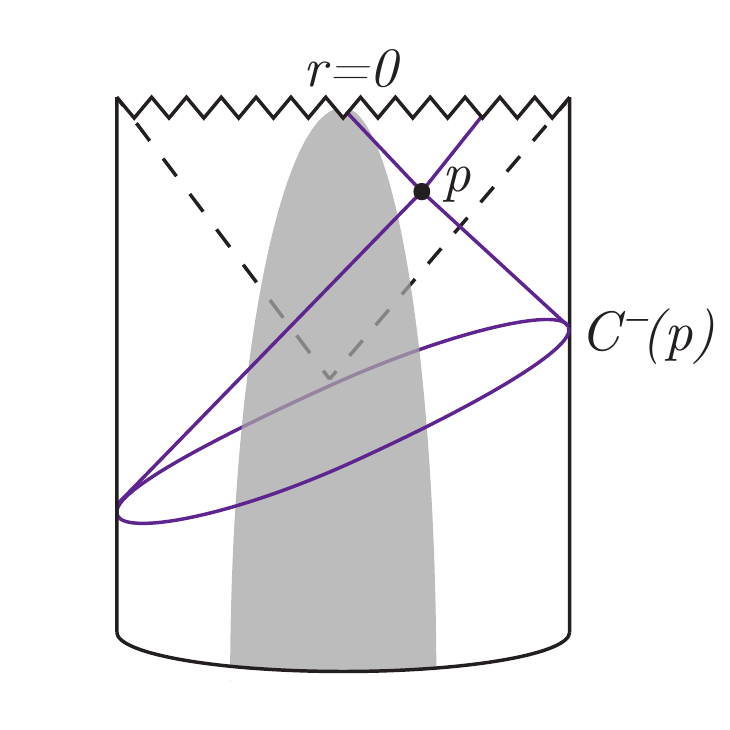}
\caption{The point $p$ lies inside the event horizon (dotted line) of a collapsing star. The boundary of the future of $p$ never intersects the asymptotic boundary. $p$ has only a past cut. More generally, the interior of an event horizon lies at least partly inside the boundary's domain of influence, but (by definition) not within the boundary's causal wedge.}
\label{fig:nofuturecutIntro}
\end{figure}

The above reconstruction of the conformal metric from the space of light-cone cuts applies to points in causal contact (either to the future or past) of the boundary. This includes points inside event horizons, and is more general than requiring that points lie in the causal wedge, which requires causal contact both to the future and past. See Fig.~\ref{fig:nofuturecutIntro} for an example.

More importantly, for (most) points inside the causal wedge, we can complete the reconstruction by 
determining the light-cone cuts without reference to the bulk. This can be done using results from~\cite{MalSimZhi} 
(which was based on earlier work by~\cite{PolSus99, GarGid09, HeePen09, Pen10, OkuPen11}),
 where it was shown how to determine a bulk point in $AdS_{d+1}$ from the singularities in
 time-ordered 
  Lorentzian  $(d+2)$-point correlators. The correlator diverges when all the boundary points are null related to a single vertex point, where the vertex can lie on the boundary or in the bulk. In cases where the vertex lies in the bulk (and there is no analogous boundary vertex) the result is a ``bulk-point'' singularity, which can be used to identify the bulk point. Extending this to  Lorentzian $(d+3)$-point correlators in excited states corresponding to  asymptotically AdS spacetimes yields a construction of light-cone cuts from the field theory. More general prescriptions may exist for finding cuts for points outside of the causal wedge that have only a past or only a future cut. This remains to be investigated.

The paper is structured as follows: in Sec.~\ref{sec:goodcut}, we define light-cone cuts more precisely,  state some of their properties, and calculate them in a simple example.  Sec.~\ref{sec:confmetric} gives (i) a way of recovering the bulk conformal metric at any bulk point in the domain of influence of the boundary from the set of light-cone cuts, and (ii) a prescription for finding the light-cone cuts associated to points within the causal wedge of the boundary. In Sec.~\ref{sec:globalrecon}, we give a partial procedure for determining the causal separation between two bulk points which need not be in some local neighborhood of one another. In Sec.~\ref{sec:discussion} we discuss possible ways of obtaining the conformal factor, $1/N$ corrections, implications for subregion/subregion duality, and possible extensions for future work.

\section{Light-cone Cuts}\label{sec:goodcut}

Recall that the chronological past of a point $p$, $I^-(p)$,  is defined to be the set of all points $q$ that can be connected to $p$ by a future-directed timelike curve. The causal past $J^-(p)$ is defined similarly, with ``timelike" replaced by ``timelike or null".
Let $M$ be an asymptotically 
AdS metric with conformal boundary $\partial M$.
We assume that $M$ is  at least $C^{2}$, maximally extended, connected  and AdS hyperbolic: there are no closed causal curves, and for any two points $(p,q)$, $J^+(p)\cap J^-(q)$ is compact after conformally compactifying the AdS boundary~\cite{Wal12}. While many of our results may be generalized to asymptotically 
AdS spacetimes with two boundaries, we will assume in this paper that $\partial M$ is connected for simplicity. All conventions, unless otherwise stated, are as in~\cite{Wald}.

The \textit{past light-cone cut} of a bulk point $p\in M$, or past cut for short, denoted $C^{-}(p)$, is defined as the intersection of the boundary of the past of $p$ with $\partial M$: $C^{-}(p)\equiv \p{J}^{-}(p)\cap \partial M$. The future cut of $p$ is defined similarly: $C^{+}(p)\equiv \p{J}^{+}(p) \cap\partial M$. See Fig.~\ref{fig:pfcutIntro} for an illustration. When a statement applies equivalently to past or future cuts, we will simply denote the cuts in question as $C(p)$. These cuts are essentially the intersection of the light cone of a bulk point with the asymptotic boundary:  a cut is not the entire light cone, however, since null geodesics can focus due to gravitational lensing. When geodesics cross, they produce caustics, which cause the null geodesic to leave the boundary of the past or future of $p$. The possible existence of caustics implies that the cuts need not be smooth cross-sections of the boundary. In general, they will be continuous, but may contain cusps where they fail to be $C^1$. We expect the cusps to be a measure zero subset of any cut, and we will assume this to be the case.

 We will now state 
three results about the correspondence between light-cone cuts and bulk points.
For pedagogical reasons, we will present our results without proof, and provide the proofs in Appendix~\ref{appendix}.  The following proposition holds for any bulk point in causal contact with the boundary (either to the future or past):\\

\noindent \textbf{Proposition:} (1) $C(p)$ is a complete spatial slice of the boundary $\partial M$, (2) For every point $p\in I^{+}[\partial M]$, there is precisely one past  cut and for every point $p\in I^{-}[\partial M]$ there is precisely one future cut,  (3)  $C(p)\cap C(q)$ contains a nonempty open set if and only if $ p=q$.\\

\noindent The proposition immediately implies that all points in the domain of influence of the boundary have at least one cut, past or future, and at most both:  points in the causal wedge have both a past and a future cut.  It establishes a one-to-one map from light-cone cuts to bulk points. For past cuts, this map covers all of $I^{+}[\partial M]$, while for future cuts it covers all of $I^{-}[\partial M]$. This  map does not always cover the entire spacetime since there may exist points without any causal contact with the boundary.

\subsection{Example: the cuts of AdS}\label{subsec:example}

In this section we provide a concrete example of a set of cuts, specifically those of AdS$_{d+1}$. For simplicity, we derive them from symmetries, although they can also be obtained by solving for the null geodesics\footnote{These cuts may also be derived from field theory data, via the prescription in Sec.~\ref{subsec:findingcuts}.}. Setting the AdS length scale to one, AdS can be obtained from the space of unit timelike vectors $P^{a}$ in a vector space with metric of signature  $(2,d)$:  $P_a P^a = -1$. This is a hyperboloid with closed timelike curves. After finding the cuts for this space, we may pass to the covering space to obtain the usual causally well behaved definition of AdS. Since the identification does not affect the relation between points and their cuts, we will refer to the hyperboloid as AdS.

The boundary at infinity is represented by null vectors $\ell^a$ in the vector space up to scaling: $\ell^a \equiv \lambda \ell^a$. The light cone of  a point $P^a$ in AdS intersects infinity at the points where $P_a \ell^a = 0$. This can be seen from the fact that if $\ell^a$ is orthogonal to $P^a$, then $P^a + s \ell^a$ with $s\in (0,\infty)$ is a curve  in AdS going from $P^a$ to the boundary. This curve is null since its tangent vector $\ell^{a}$ is null. Thus the cut of a point $P^a$  just consists of the orthogonal null vectors.

To make this more explicit, we introduce coordinates $(T_1, T_2, X^i)$ with $i = 1, \cdots, d$ so AdS is given by 

\begin{equation}\label{eq:AdS} 
-T_{1}^{2}-T_{2}^{2} +X_{i} X^{i}=-1,
\end{equation}
 These coordinates are related to AdS global coordinates via the following map:
\begin{equation} r^{2}=X_{i}X^{i} , \ \ T_{1}= \sqrt{r^{2}+1} \ \sin t, \ \ T_{2}= \sqrt{r^{2}+1}\ \cos t  \label{eq:globalcoord}
\end{equation}
\noindent where the angular coordinates transform in the obvious way\footnote{The AdS metric in these coordinates takes the familiar form:
\begin{equation}ds^{2} = -(r^{2}+1)dt^{2} +(r^{2}+1)^{-1}dr^{2} + r^{2}d\Omega^{2}.
\end{equation}}.
We will take the boundary at $r\rightarrow \infty$ to be in the conformal frame of the Einstein Static Universe, or the static cylinder,  with metric:
\begin{equation} \label{eq:ESU} ds^{2}=-dt^{2} + d\Omega^{2},
\end{equation}
\noindent where the time and angle coordinates are the same as those of AdS.  

To connect this with the abstract definition of the cut, let $T_1^a, T_2^a, X_i^a$ be an orthonormal basis corresponding to the above coordinates. To begin, suppose $P^a = \cos t_0 \ T_2^a + \sin t_0\ T^a_1$. This corresponds to the point $t=t_0, r=0$ in AdS. The orthogonal timelike direction is $\xi^a = -\sin t_0\ T_2^a + \cos t_0\ T^a_1$, so $\ell^a$ is the sum of $\xi^a$ and any unit vector in the $X_i^a$ directions. To describe the cut, we note from \eqref{eq:globalcoord} that $\tan t_{\infty} = T_1/T_2$, where $t_{\infty}$ is the value of the time coordinate $t$ at the cut. Since $T_1$ and $T_2$ are just the coefficients of the corresponding basis vectors in $\ell^a$ we have 
\be
\tan t_{\infty} = \frac{T_1}{T_2} = - \cot t_0
\ee
So the cut is simply given by $t_{\infty} = t_0 \pm \pi/2$, where the plus sign refers to future cuts and the minus sign refers to past cuts. Of course this result could have been obtained just from spherical symmetry and the fact that light rays take $\pi/2$ coordinate time to get to the boundary. The advantage of this approach is that it yields the cuts for points off the axis just as easily.

To see this, let $P^a = \sqrt{1 + r_0^2} \ T_2^a + r_0\ X_1^a$. This corresponds to a point at $t_0=0$ and $r=r_0$ in the direction $X_1$. To find the orthogonal null vectors, we expand $\ell^a = c_1 T_1^a + c_2 T_2^a + d_i X_i^a$. Imposing $P\cdot \ell = 0$ and $\ell\cdot \ell = 0$, and solving for $\ell^{a}$, we obtain
\begin{equation} \label{eq:AdScuts}
\tan t_{\infty}(\theta) = \frac{c_1}{c_2} =  \frac{\sqrt{1+r_0^{2}\sin^{2}\theta}}{r_0 \cos\theta}.
\end{equation}
where $\theta$ is the angle  with the $X^{1}$ axis. This cut is tilted with respect to a cut at constant $t_{\infty}$. The cut for an arbitrary point can be obtained from this one by time translations and rotations,  so there is a $(d+1)$-dimensional space of cuts, labeled by $t_0, r_0,$ and the $(d-1)$-directions of the tilt.
Since there are no caustics in AdS, these cuts are all smooth. Note that in the limit $r_0\to \infty$ (which corresponds to the bulk point approaching the boundary), the cut reduces to $t_\infty= \pm \theta$; this is just a null cross-section of the boundary. 

For pure AdS, there is a simple relation between the behavior of the cuts and the global causal relation of the corresponding points\footnote{We say that two points are spacelike separated if there exists no causal curve between them; null separated if there exists a null achronal curve between them but no timelike curve between them; timelike separated if there exists a timelike curve between them.}: $C(p)$ and $C(q)$  do not intersect if and only if $p$ and $q$ are timelike related, $C(p)$ and $C(q)$ intersect at precisely one point if and only if they are null related, $C(p)$ and $C(q)$ 
intersect at more than one point if and only if $p$ and $q$ are spacelike related. These results are intuitively clear and follow from the results in the appendix (and the fact that there are no caustics in AdS). We will see that only some of these results extend to general asymptotically AdS spacetimes.

There is an intrinsic way of characterizing these light-cone cuts of AdS. The light cone of a point in AdS is shear-free. Given any smooth spacelike cross-section of the boundary, the shear of the congruence of ingoing orthogonal null geodesics from that cross-section always vanishes asymptotically as ${\cal O}(1/r^2)$. Demanding that the $r^{-2}$ contribution to the shear vanishes yields a differential equation for the cross-sections with solutions given precisely by \eqref{eq:AdScuts}. In the case of asymptotically flat spacetimes, this was the starting point for H-space: the set of asymptotically shear-free cuts of null infinity (see e.g.~\cite{New76, HanNewPen} and \cite{HSpaceRev} for a review). However, this characterization is less useful for us, since in generic asymptotically AdS spacetimes, light-cone cuts are not asymptotically shear-free in this sense. We will discuss another intrinsic way to characterize the light-cone cuts in the next section; instead of geometric constructs such as shear, we will make use of a tool not available to~\cite{New76, HanNewPen}: the dual field theory.


\section{Reconstruction of the Conformal Metric}\label{sec:confmetric}

In this section we will first consider bulk points in the past or future of the conformal boundary and show how to reconstruct the bulk conformal metric given the set of cuts. We can work with either past or future cuts, but will focus on past cuts for definiteness. We will then describe a way of constructing the light-cone cuts associated with (most) points in the causal wedge from field theory data. There are indications that  more general procedures may exist. 

\subsection{Reconstructing the conformal metric from the cuts} \label{subsec:reconcausal}
Under the assumption that we have been provided with the set of light-cone cuts, past or future, we would like to determine the conformal metric. Below we do this in two steps: the first is a result showing that the conformal metric at a point is fixed by any open subset of the point's light cone; the second is a prescription, making use of a result proven in the appendix, for constructing the conformal metric from the set of cut locations.

The conformal metric at a point is simply the metric up to an overall positive constant: $g_{\mu\nu} \equiv \lambda^2 g_{\mu\nu}$. Clearly, two conformally related metrics have identical light cones. Conversely, the conformal metric at a point is uniquely fixed by the light cone at that point. Since an open set of a cut fixes the entire cut, an even stronger result is true: the conformal metric is uniquely fixed by any open subset of the light cone. This result was proven in~\cite{KozNew83}; here we give a different argument which will be useful below  in reconstructing the  metric from the cuts.

In $d+1$ dimensions, we may take any $d+1$ linearly independent past- or future-pointing null vectors $\ell_{i}$ at a point $p$, and view them as a basis of the tangent space at $p$. This is always possible, since there is no vector orthogonal to all of the null vectors. The vectors $\ell_i$ may lie anywhere on the lightcone of $p$: we need only an open subset of the lightcone to find this basis. By definition, the $\ell_{i}$ all have zero norm, but unknown inner products; the conformal metric at $p$ is precisely fixed by these inner products. 
  To determine them, take a new collection of null vectors, $\eta_k$, and expand them in terms of $\ell_i$: 
\begin{equation} \eta_{k} = \sum\limits_{i} M_{ki}\ell_{i}.
\end{equation} 
Each $\eta_{k}$ has zero norm by definition; this yields a set of algebraic equations for the inner products $\ell_i \cdot \ell_j$:
\begin{equation}0=  \eta_{k}\cdot \eta_{k} = \sum\limits_{i,j} M_{ki}M_{kj} (\ell_{i}\cdot \ell_{j}) \qquad {\rm no\ sum\ on \ }k.
\end{equation}
While it is not generally true that such equations must have a solution, we are guaranteed a solution precisely because these equations describe a Lorentzian metric which by construction exists. By choosing enough vectors $\eta_k$, we will find a solution which is unique up to an overall constant rescaling of all inner products. This determines the conformal metric at $p$.\footnote{Repeating this construction at each point yields a smooth tensor field, which in particular includes complete information about all of its derivatives.}

We will now implement this approach to recover the conformal metric from the cuts. Suppose that we are given the set of past cuts ${\cal M}$. From the proposition, this is a $(d+1)$-dimensional space representing all bulk points in $I^+[\p M]$. We now define a conformal metric on ${\cal M}$. To do so, we  use the following result (proven in the Appendix): \\

\noindent \textbf{Theorem 1:} If $C(p)$ and $C(q)$ intersect at precisely one point, and both cuts are $C^{1}$ at this point, then $p$ and $q$ are null-separated.\\

\noindent The crux of the proof is in the uniqueness of the inward-directed orthogonal null geodesics $\gamma$ from every $C^1$ point of $C(p)$. If $C(p)$ and $C(q)$ are tangent at a regular point of both cuts, $\gamma$ must lie on the boundary of both $J^-(p)$ and $J^-(q)$. This is only possible if $\gamma$ goes through both $p$ and $q$, so the points $p$ and $q$ must be null-separated. The result proved in the Appendix is actually stronger, and shows that there exists a cut tangent to $C(p)$ for every bulk point  along an achronal null geodesic from $p$ to the boundary.

\begin{figure}[t]
\centering
\includegraphics[width=8cm]{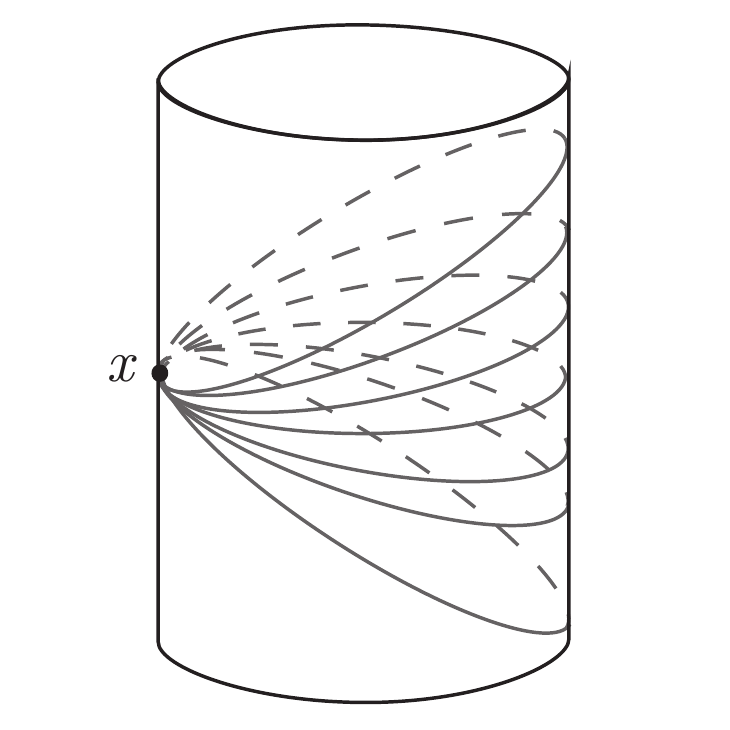}
\caption{Cuts corresponding to a null bulk geodesic. The cuts are all tangent at the point $x$ at which the null geodesic reaches the boundary.}
\label{fig:nullgeodesic}
\end{figure}

\begin{figure}[t]
\centering
\subfigure[]{
\includegraphics[width=6cm]{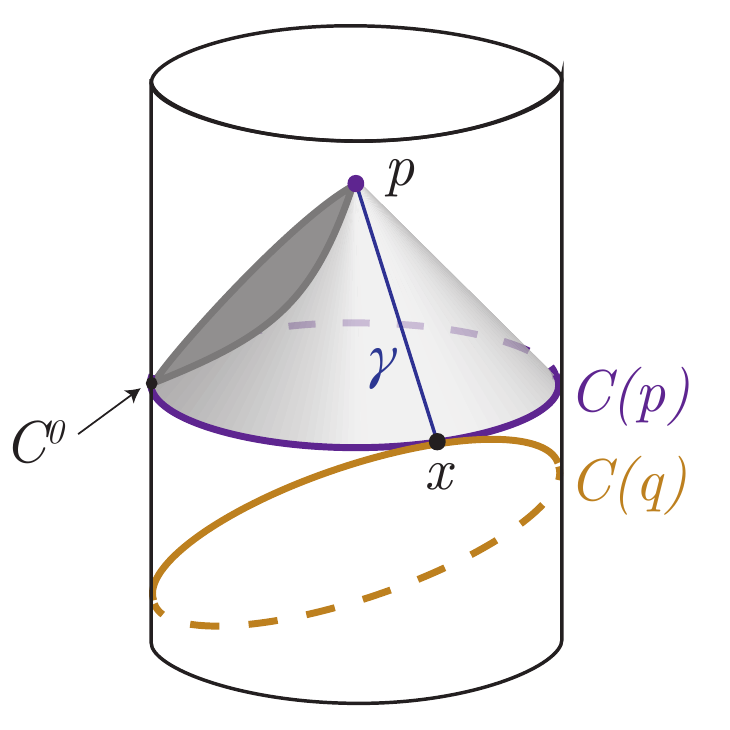}
\label{subfig:crossing}
}
\hspace{1cm}
\subfigure[]{
\includegraphics[width=6cm]{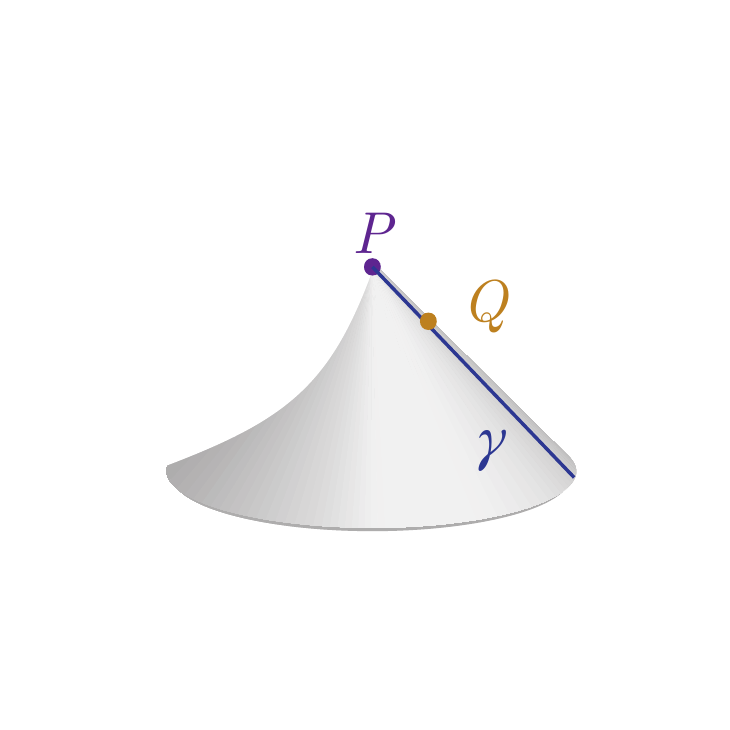}
\label{subfig:sandwich}
}
\caption{(a): $\partial J^{-}(p)$ will generally have caustics and some isolated $C^{0}$ points on the cut $C(p)$. At any regular point $x$, there is a null achronal geodesic $\gamma$ from $p$ all the way to $C(p)$. 
(b): In the space of cuts ${\cal M}$, a point $P$ corresponds to a cut $C(p)$; the null curve $\gamma$ of $\partial J^{-}(p)$ corresponds to a null curve $\gamma$, where points $Q$ on $\gamma$ are cuts $C(q)$ which are tangent to $C(p)$ at $x$.}
\label{fig:spaceofcuts}
\end{figure}

Theorem 1 endows the space of cuts with a natural  Lorentzian structure, inherited from the bulk Lorentzian structure: given a point $P$ in ${\cal M}$, i.e., a cut $C(p)$, the set of all other cuts which are tangent to $C(p)$ at a regular point $x$ forms a null curve in ${\cal M}$; this null curve precisely corresponds to the unique null bulk generator shared by all the cuts which are tangent at the regular point $x$. See Fig.~\ref{fig:nullgeodesic} for an illustration. More generally, the set of all cuts which are tangent to $C(p)$ at any regular point forms a null hypersurface in ${\cal M}$. The tangent vectors to this null hypersurface at $P$ form (a part of) the light cone of $P$, see Fig.~\ref{fig:spaceofcuts}, just as the unique null generators fired from all regular points of $C(p)$ form a subset of the bulk lightcone of $p$. To reconstruct the bulk conformal metric at $p$, we need only recover it at $P$.

By the reasoning above, a set of $d+1$ regular points of $C(p)$, with cuts tangent to each of the regular points,  maps to a set of $d+1$ linearly independent null vectors on the lightcone of $P$. As argued in the beginning of this section, this uniquely determines the conformal metric at $P$, and therefore immediately also the conformal metric at 
$p$\footnote{Note that generically, the entire lightcone cannot be recovered in this way due to $C^0$ points on the cut arising from caustics in the bulk, but that is not required to fix the conformal metric at $P$.}; the additional null vectors $\eta_{k}$ required to determine the conformal metric at $P$ may be obtained from cuts which are tangent to $C(p)$ at other points.

We emphasize that given the past cuts, we can recover the conformal metric at all points in $I^+[\p M]$ this way, even points inside black holes. The fact that $q$ might be on a caustic of $\p I^-(p)$ for some point $p$ is not an obstacle to finding the conformal metric at $q$.

\subsection{Finding the light-cone cuts}\label{subsec:findingcuts}

The distinguishing characteristic of a cut $C(p)$ is that every point on $C(p)$ is null-related to the same point $p$ in the bulk. More generally, $C^{-}(p)$ and $C^{+}(p)$ are the past and future cuts of a bulk point $p$ whenever there exist null geodesics from every point on $C^{-}(p), C^{+}(p)$ to $p$. 

A similar structure was used recently to identify a bulk point in $AdS_{d+1}$  from $(d+2)$-point
time-ordered
 Lorentzian correlation functions in the dual field theory~\cite{MalSimZhi}. In general, $n$-point functions have divergences when all points are null separated from an interaction vertex and energy-momentum conservation holds at the vertex. In~\cite{MalSimZhi}, it was shown that there are cases where the correlators diverge due to an interaction vertex in the bulk that was null separated from the boundary points,  but there is no analogous vertex point on the boundary to explain the divergence. Such singularities were termed ``bulk-point singularities''.

These bulk-point singularities can be used to uniquely specify a bulk point. In $d+1$ bulk dimensions, there is a $d$-dimensional subspace of points which are connected to a point on the boundary by a future-directed null geodesic. Given $d+1$ boundary points, then, these null subspaces can intersect at most at a point. Conservation of energy momentum at the vertex requires one more boundary point, so singularities of $(d+2)$-point correlators can be used to fix a bulk point. Singularities of higher point correlators can also be used to fix bulk points.

Although the analysis in~\cite{MalSimZhi} was restricted to vacuum correlation functions dual to pure AdS, similar behavior should occur for correlation functions in excited states corresponding to asymptotically AdS spacetimes \cite{MalPC}. In fact, the generalization to asymptotically AdS spacetimes has an advantage. In higher dimensional $AdS_{d+1}$, it can be difficult to show that the singularity in the correlator is in fact due to a bulk-point singularity and not an ordinary null field theory singularity. It is crucial for our construction to work that the singularity be sourced by a bulk null separation. To construct bulk points from correlation function singularities, we must show that there is no boundary point which is null related to all $d+2$ points in the correlation function.  In \cite{MalSimZhi}, this was shown  for $d = 2,3$. This is, in fact, simpler to show in any dimension away from pure AdS: when the spacetime is not exactly AdS, null geodesics take longer to pass through the bulk than they do on the boundary~\cite{GaoWal00}. Thus, the bulk-point singularities away from AdS will occur when the boundary points are separated in time by more than the light travel time on the boundary, which immediately precludes the possibility of a null related boundary vertex point\footnote{This relies on gravitational delay in the bulk, which is true under assumption of the Averaged Null Curvature Condition, a condition expected to be obeyed by low energy supergravity limits of string theory.}.

\begin{figure}[t]
\centering
\includegraphics[width=8cm]{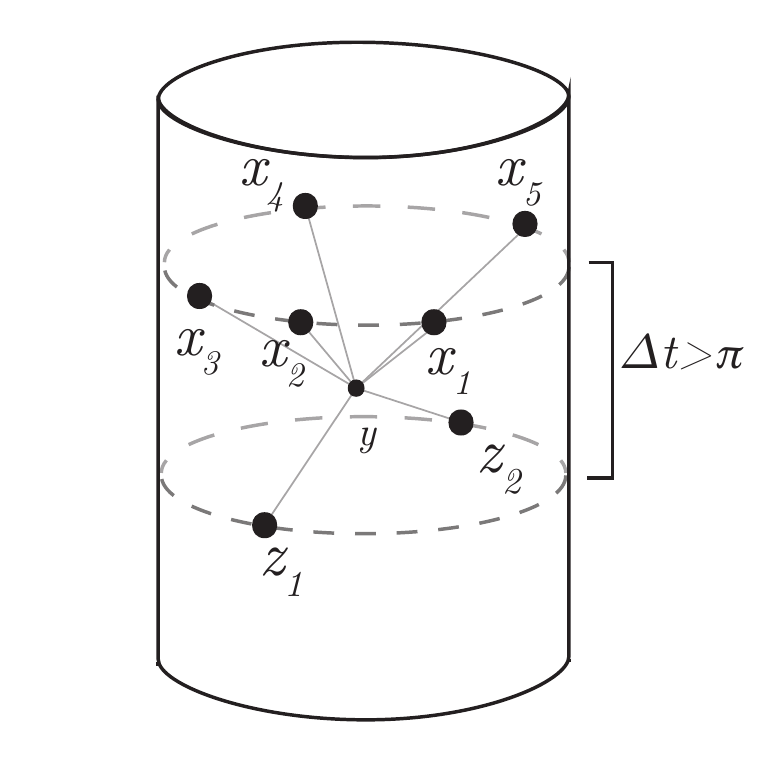}
\caption{A Landau diagram of a bulk-point singularity in a 7-point function: $z_1$, $z_2$, and the $x_{i}$ are all null-separated from a bulk point $y$ so that high energy test particles from $z_1, z_2$ scatter at $y$, conserving energy and momentum. To find the past cut of $y$, we vary $z_1, z_2$ in a spatial direction while keeping the 7-point function singular.}
\label{fig:findingcuts2}
\end{figure}

Using these bulk-point singularities, we can identify the past and future cuts of most bulk points in the causal wedge.  To determine the cut, we want to move some of the boundary points while keeping the vertex point fixed in the bulk. In order to keep energy momentum conserved at the vertex we need one more boundary point. So we start by taking two boundary points $z_1$ and $z_2$ which are spacelike related to one another and $d+1$ other boundary points $x_{1},\cdots ,x_{d+1}$ where the $x_{i}$ are to the future of  $z_{1}, z_{2}$  and  spacelike related to each other (see Fig.~\ref{fig:findingcuts2}). Consider deforming the set of points $(z_1, z_2, x_j)$ until the $(d+3)$-point  correlation function diverges: 
\begin{equation} \label{eq:divergence}
\left \langle \mathcal{O}(z_1)\mathcal{O}(z_2) \mathcal{O}(x_{1})\cdots \mathcal{O}(x_{d+1})\right\rangle\rightarrow \infty, 
\end{equation}
 The divergence in Eq.~\ref{eq:divergence}  will occur whenever the points $(z_1, z_2, x_j)$ are all null-related to a vertex $y$  in the bulk or the boundary, and high energy test quanta fired from $z_1, z_2$  scatter at $y$ to the $x_{i}$, where energy and momentum are conserved at $y$. In terms of the global time on the static cylinder,  any two points on the boundary which are null separated must have time separation less than $\pi$. By taking $x_i$ to be more than a time $\pi$ to the future of $z_1$ and $z_2$, we can ensure that they will not be null related to a point on the boundary. 
 Note that there cannot be more than one vertex in the bulk:  energy momentum conservation requires at least two incoming  and two outgoing quanta at each vertex, and there are only two incoming quanta.

We now fix the points ${x_i}$, which fixes a point $y$ in the bulk, and vary
 $z_1$ and $z_2$ (requiring that they remain on a spacelike boundary slice). See Fig.~\ref{fig:findingcuts2} for an illustration. The collection of all points $z_1, z_2$ satisfying Eq.~\ref{eq:divergence} will trace out  the cut $C^{-}(y)$. Here, however, we must issue a caveat: since energy and momentum must be conserved at the vertex, we may not recover the entire cut this way. Generically, the existence of caustics means that only part of the light cone of $y$ is connected to $C^{-}(y)$ by null geodesics. Let us call this subset of the light cone $N$. We can recover parts of the cut corresponding to pairs of points in $N$ whenever energy and momentum can be conserved at $y$.\footnote{It is possible that more of the cut can be recovered by considering singularities in higher point correlators.}  Fortunately, this restriction does not affect our bulk reconstruction. Since we are fixing the $d+1$ future points, we still get a one-to-one map between our partial cuts and bulk points. Furthermore, we can still construct the nearby cuts that are tangent at $C^1$ points, which determines the conformal metric at that point.  
 By reversing the above construction, we may similarly construct the future cut $C^{+}(y)$.

If $N$ is too small,  then we cannot recover any of the cut, and our prescription for obtaining cuts will fail. This happens, for example, close to a black hole. Consider a 
static
spherical AdS black hole for simplicity, and let $r_n$ denote the radius of the closed null geodesic around the black hole. Then for $r<r_n$,  most of the light rays fall into the black hole. Less than half the light cone makes it out to the boundary, and our construction is insufficient.  For $r $ slightly larger than $ r_n$, more than half of the light cone makes it out to $\p M$,  but not all of the null geodesics stay on the boundary of $J^-(p)$ and reach $C(p)$. This should not be a problem since we expect singularities in the correlators for all boundary points which are related to the vertex by a null geodesic, even if the null geodesic is not achronal. Thus we should be able to recover part of the cut for all $r > r_n$. However, points with $r<r_n$ constitute a  ``shadow region" around a black hole where we cannot recover the cuts from field theory correlators in this way.

The divergence in the correlation function should be present in the large $N$ and large $\lambda$ limit of any holographic field theory  (potentially including perturbative corrections  in $1/N$ and $1/\lambda$). These divergences are expected to disappear at finite $N$ and $\lambda$, in agreement with the intuition that nonperturbative stringy and quantum effects fuzz out bulk points.

We emphasize that we have described just one way of obtaining the cuts from field theory data; other procedures may exist. If the bulk has certain symmetries, for instance, there will be a distinguished set of cuts that correspond to fixed points of these symmetries.   As an example, consider the field theory dual to a spherically symmetric collapse in the bulk. The dual field theory undergoes thermalization, and the field theory stress tensor is spherically symmetric.  There are preferred cuts on the boundary which are invariant under this spherical symmetry. These cuts are precisely the past cuts of bulk points at the origin, which are fixed points of the symmetry. Such cuts include points that lie inside the black hole event horizon. This approach admittedly does not yield a complete reconstruction of the conformal metric inside the black hole interior, and it requires the strong assumption of spherical symmetry, but it indicates that there may be additional ways of obtaining cuts.

\section{Some Global Causal Relations from Cuts}\label{sec:globalrecon}

The set of cuts contains more information about the causal relations between bulk points than we used in Sec.~\ref{subsec:reconcausal}, where the focus was on null separations.  In this section we describe some further results. We will consider bulk points in the causal wedge of the entire boundary, and assume that we are given the complete set of past and future cuts of these points. We note that some of the results below only use one set of cuts; those results hold everywhere within the boundary's domain of influence. 

Let $\{C^{+}(p), C^{-}(p)\}, \ \{C^{+}(q), C^{-}(q)\}$ be two distinct pairs of cuts, with corresponding bulk points $p$, $q$. We start with a definition:\\

\noindent \textbf{Definition:} $C(p)$ and $C(q)$ {\it cross} 
if $C(p)\cap I^{+}(C(q))\neq \varnothing$ and $C(q)\cap I^{+}(C(p))\neq \varnothing$.\\

\noindent This is the case when the the intersection $C(p)\cap C(q)$ divides $C(p)$ and $C(q)$ each into two or more connected (nonempty) components.

The following result (which is proved in the Appendix) tells us $p$ and $q$ are spacelike separated under the following conditions:\\
\begin{figure}[t!]
\centering
\subfigure[]{
\includegraphics[width=5cm]{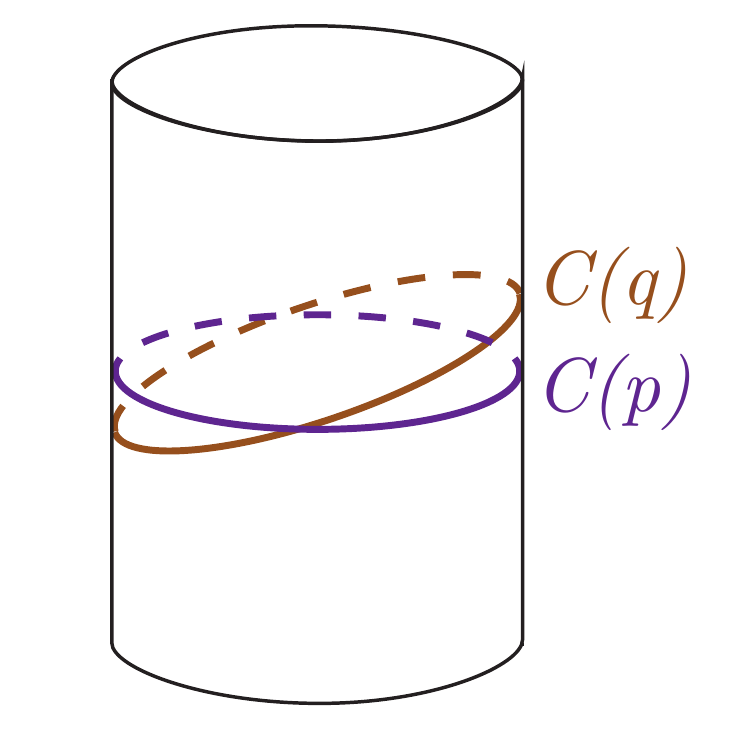}
\label{subfig:crossing}
}
\hspace{1cm}
\subfigure[]{
\includegraphics[width=5cm]{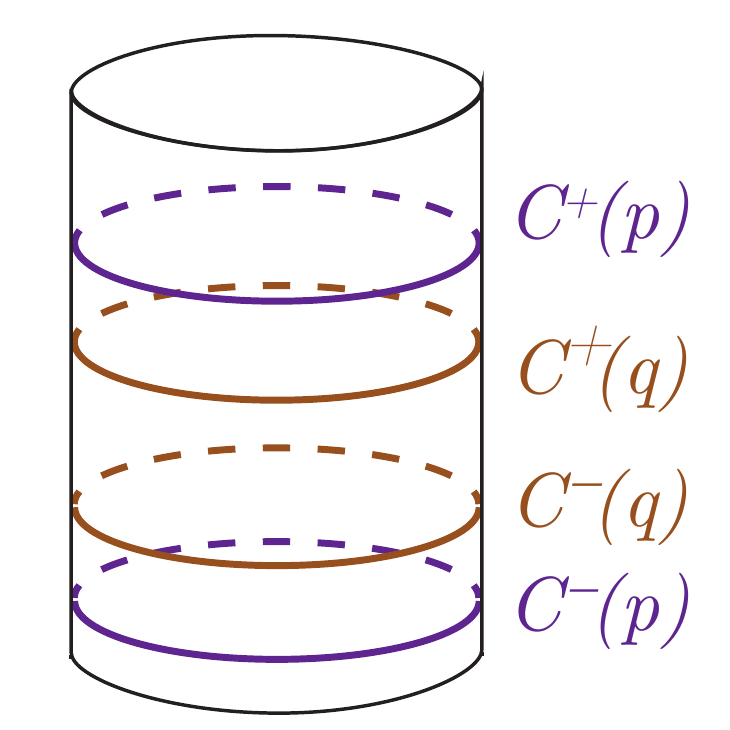}
\label{subfig:sandwich}
}
\caption{Two points $p$ and $q$ are spacelike separated if \subref{subfig:crossing} their cuts cross, or  \subref{subfig:sandwich} $C^\pm(q)$ both lie between $C^+(p)$ and $C^-(p)$.}
\label{fig:spacelike}
\end{figure}
\begin{figure}[t!]
\centering
\includegraphics[width=5.5cm]{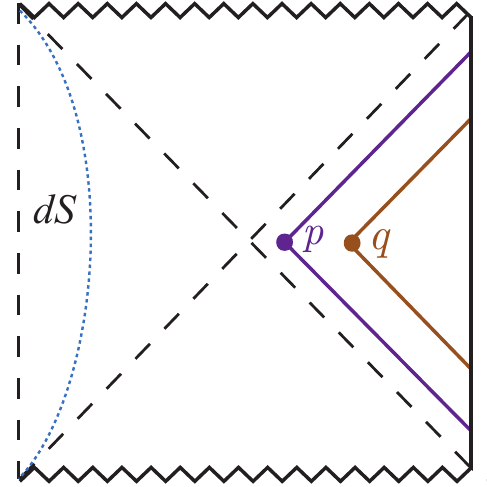}
\caption{A bag of gold geometry, where the region behind the horizon contains part of de Sitter space~\cite{FreHub05}. Points in the dS region are in the causal shadow: they have no corresponding boundary cuts. Within the causal wedge, points that are spacelike separated at some large distance will feature the more unusual  cut configuration of Fig.~\ref{subfig:sandwich}.}
\label{fig:bagofgold}
\end{figure}

\noindent \textbf{Theorem 2:} $p$ and $q$ are spacelike separated if  one of the following is true:

\begin{enumerate}
	\item  $C(p)$ and $C(q)$ cross, where $C(p)$ and $C(q)$ are either both past or both future cuts (Fig.~\ref{subfig:crossing}).
	\item $C^{\pm}(q)$ both lie between $C^{+}(p)$ and $C^{-}(p)$ (Fig.~\ref{subfig:sandwich}).
\end{enumerate}

\begin{figure}[t]
\centering
\includegraphics[width=5cm]{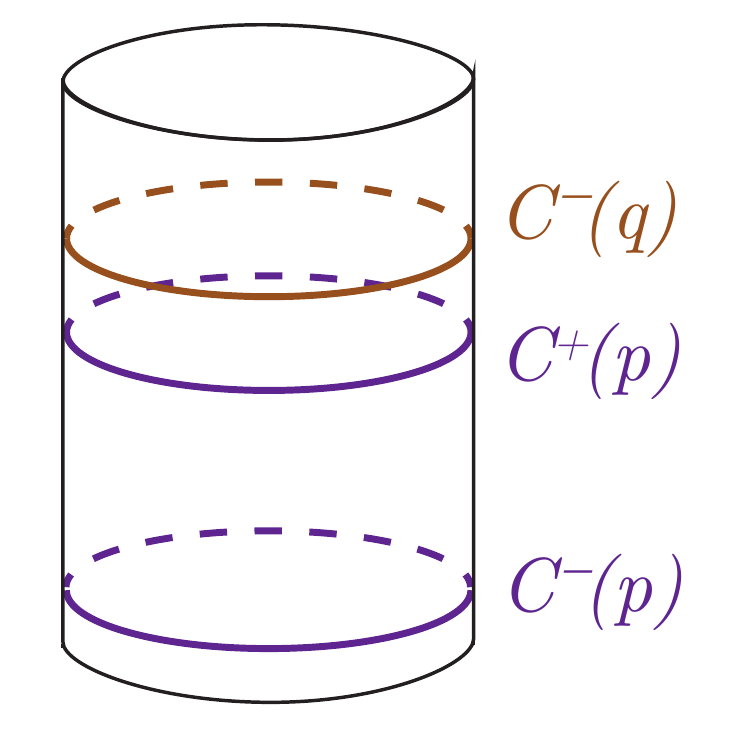}
\caption{A point $q$ is in the future of $p$ whenever $C^-(q)$ lies to the future of $C^+(p)$.}
\label{fig:timelike}
\end{figure} 

\noindent Case (1) is in agreement with expectation: we expect that two points are spacelike-separated if  their pasts or futures intersect, but are not proper subsets of one another.  Case (2) is more unusual, but can arise e.g. when one point is close to a bifurcation surface of a black hole. Consider the bag of gold geometry shown in Fig.~\ref{fig:bagofgold}, as  constructed in~\cite{FreHub05}: $p$ clearly has future-directed outgoing null geodesics which reach the boundary at late time. Since the cut must be spacelike, the entire future cut $C^+(p)$ exists at late time. Note that while some null geodesics enter the black hole and do not make it out to infinity, there are others which curve around the horizon and reach the boundary, so $C^+(p)$ is always a complete spacelike slice of $\p M$, as required by the 
proposition
in section 2. Similarly,   there are past-directed null geodesics which reach the boundary at early time, so $C^-(p)$ exists at early time. Any point farther from the horizon, like $q$ in the figure, will have $C^{\pm}(q)$ both lying between $C^{+}(p)$ and $C^{-}(p)$.

We now consider timelike separated points in the bulk. There is a simple condition on the cuts which ensures that the bulk points are timelike separated: there is a future-directed timelike path from $p$ to $q$ ($q\in I^{+}(p)$) if $C^{-}(q)$ is in the (chronological) future of $C^{+}(p)$, see Fig~\ref{fig:timelike}. This follows since there is a past-directed causal path from $q$ to $C^{-}(q)$, another from $C^{-}(q)$ to $C^{+}(p)$, and another from $C^{+}(p)$ to $p$. Since this causal path contains a timelike segment from $C^{-}(q)$ to $C^{+}(p)$, $p$ and $q$  are timelike related. This condition is limited to points separated by a sufficiently long time. 

\begin{figure}[t]
\centering
\includegraphics[width=6cm]{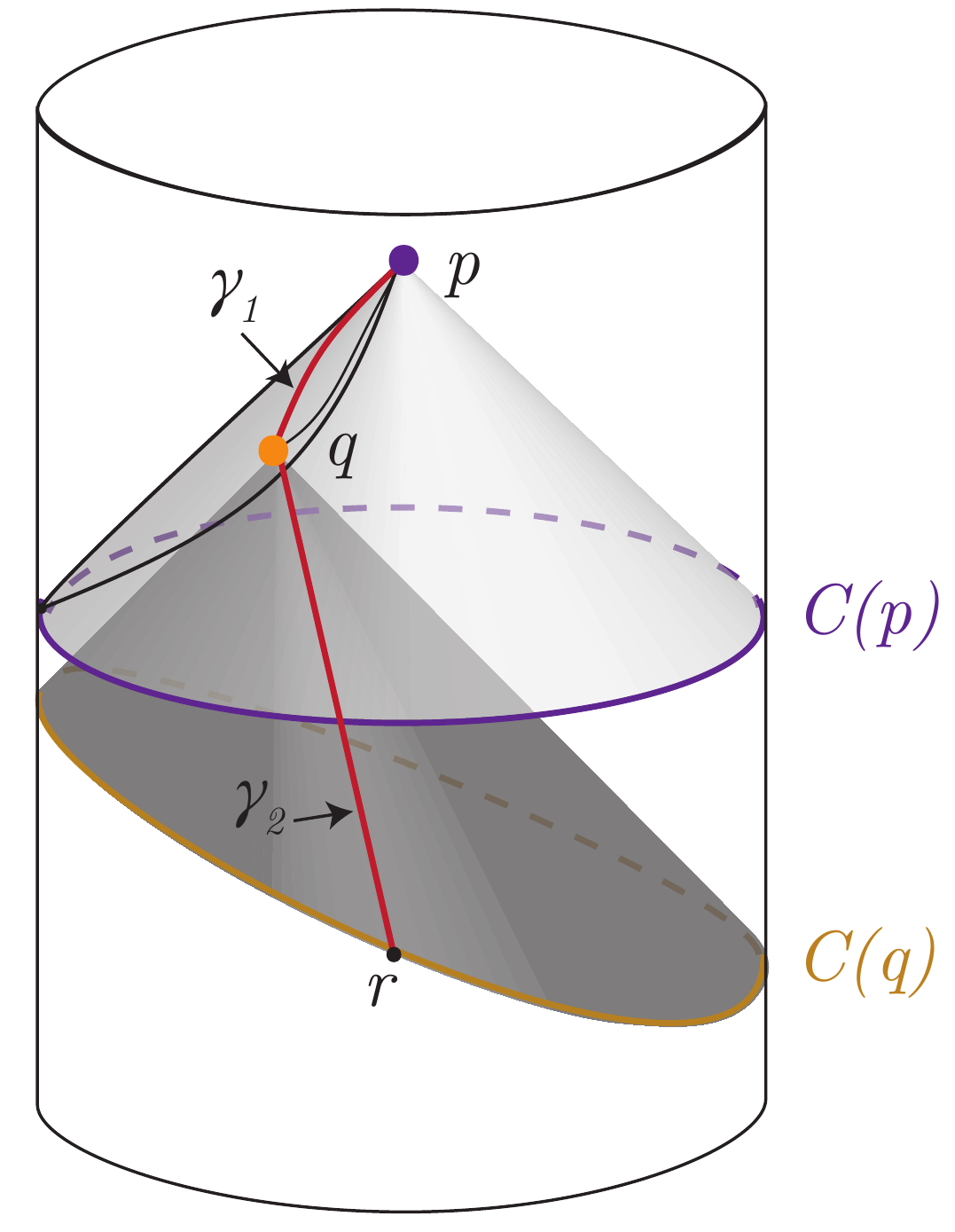}
\caption{When there are caustics, $C(q)$ can lie in the past of $C(p)$ even though $p$ and $q$ are null related.}
\label{fig:Caustics}
\end{figure}

To determine when two  points are timelike separated with shorter time differences is more difficult. It is true that if $q\in I^{-}(p)$, then $C^-(q) \subset I^{-}[C^-(p)]$ and the cuts do not intersect. The converse, however, is false: if $\p J^-(p)$ contains a line of caustics, and $q$ is a point on this line, then there is a null geodesic $\gamma_1$ from $p$ to $q$. These two points are therefore null related. By definition, for any point $r \in C^-(q)$, there is a null geodesic $\gamma_2$ from $q$ to $r$. Combining $\gamma_{2}$ with $\gamma_1$, we obtain a broken null geodesic from $p$ to $r$; this broken null geodesic can always be lengthened by rounding out the corner\footnote{If the two geodesics happen to join smoothly without a corner, $r$ lies on the continuation of $\gamma_1$. Since this geodesic encounters a caustic, $r$ must still be in the past of $p$.}. 
This means that the entire cut $C^-(q)$ is in the past of $p$, and  $C^-(q) \subset I^{-}[C^-(p)]$. See Fig.~\ref{fig:Caustics}. Since these two cuts do not intersect even though the points are null related, we can move $q$ slightly to the future or past and get nonintersecting cuts for spacelike, timelike or null separated points. 

Finally, we note that some spacetimes have  points which have no causal contact at all with the boundary (e.g. points  inside the bag of gold geometry~\cite{FreHub05} shown in Fig.~\ref{fig:bagofgold}, or points deep in the throat of the shockwave geometries of~\cite{SheSta14}). It is not clear at this time if there is any generalization of the notion of a light-cone cut that would apply to these points.

\section{Discussion}\label{sec:discussion}

We have presented a new approach for reconstructing the bulk spacetime from its dual field theory. Given a $d$-dimensional field theory, there is a $(d+1)$-dimensional space of light-cone cuts which represent the points of the holographic spacetime.  The cuts can be determined from singularities in correlation functions, and the bulk metric (up to a conformal rescaling) is simply recovered from the location of the cuts. This procedure works for most points in both past and future causal contact with the boundary, i.e. the causal wedge of the entire boundary. (There is a ``shadow region" around e.g. a 
static black hole where our procedure fails to determine the cuts.)

There are many open questions related to possible extensions of these results.  We have  shown that the conformal metric can be recovered from cuts associated with any points that are within the domain of influence of the boundary. The problem now is to find new ways to recover the cuts from field theory data. We would like to determine cuts in the shadow region around black holes as well as points outside the causal wedge. As discussed in Sec.~\ref{subsec:findingcuts}, symmetries can facilitate finding such cuts. This suggests that there may be a more general prescription for obtaining the light-cone cuts in such spacetimes, in keeping with recent arguments that bulk reconstruction is possible beyond the causal wedge, all the way to the entanglement wedge~\cite{CzeKar12, Wal12, HeaHub14, JafSuh14, JafLew15, DonHar16}. 

Another important extension is the determination of the conformal factor from boundary data. 
In cases when the conformal factor is analytic, we may easily obtain the conformal factor from a Fefferman-Graham expansion near the boundary, where the free coefficients are fixed by field theory expectation values. In general, any analytic spacetime metric can be obtained in this way, but we emphasize that this construction is more general: the conformal metric need not be analytic; only the conformal factor does. 
 
We have focused on the metric, but it is also interesting to ask whether other 
bulk fields can be recovered in a natural way from light-cone cuts. One possible direction to explore is suggested by the
 similarity between our construction and twistor theory. The latter involves considering the space of all null geodesics in Minkowski space, and identifying a spacetime point by the sphere of null geodesics passing through it. Spacetime fields are then encoded in certain singular holomorphic functions on (complexified) twistor space. Possibly some equally elegant prescription for describing spacetime fields exists on the space of light-cone cuts. 

Other extensions include the generalization to spacetimes where the boundary is not connected, such as the two-sided black hole.  It is plausible that many of the proofs would carry over to such cases, but we have not yet shown this rigorously. 
There are also hints of connections between light-cone cuts and bulk singularities, which should be explored further.  The existence of past cuts which have no corresponding future cut (or vice versa) is an indication that some future-directed (past-directed) null geodesics never reach the asymptotic boundary. Generically this indicates null geodesic incompleteness in the bulk, although exceptions may exist (e.g. null geodesics trapped in orbit).

We emphasize that our construction is entirely covariant, and that proofs of the necessary results rely exclusively on causal structure and continuity arguments. No assumptions have been made regarding the bulk equations of motion or matter content besides those which follow from field theory causality. This naturally raises the question of which, if any, of our results remain valid when the bulk undergoes quantum and stringy corrections.  The  reconstruction procedure remains valid so long as there is a well-defined notion of bulk causal structure. 
 This is true even with perturbative quantum ($1/N$ ) or stringy ($1/\lambda$) corrections. The work of~\cite{MalSimZhi} suggests that bulk-point singularities exist perturbatively, so the procedure of Sec.~\ref{subsec:findingcuts} should also work when   perturbative corrections are included. The result would be the expectation value of the conformal metric, rather than a metric operator.

Nonperturbative quantum physics in the bulk of course remains mysterious, and it is not clear that there is a good notion of causality. Certainly we do not expect that there is a sharply defined notion of points and distances. In particular,~\cite{MalSimZhi} showed that bulk-point singularities vanish nonperturbatively, so the correlation function prescription of Sec.~\ref{subsec:findingcuts}  will no longer work at finite $\lambda, N$. 

\begin{figure}[t]
\centering
\includegraphics[width=8cm]{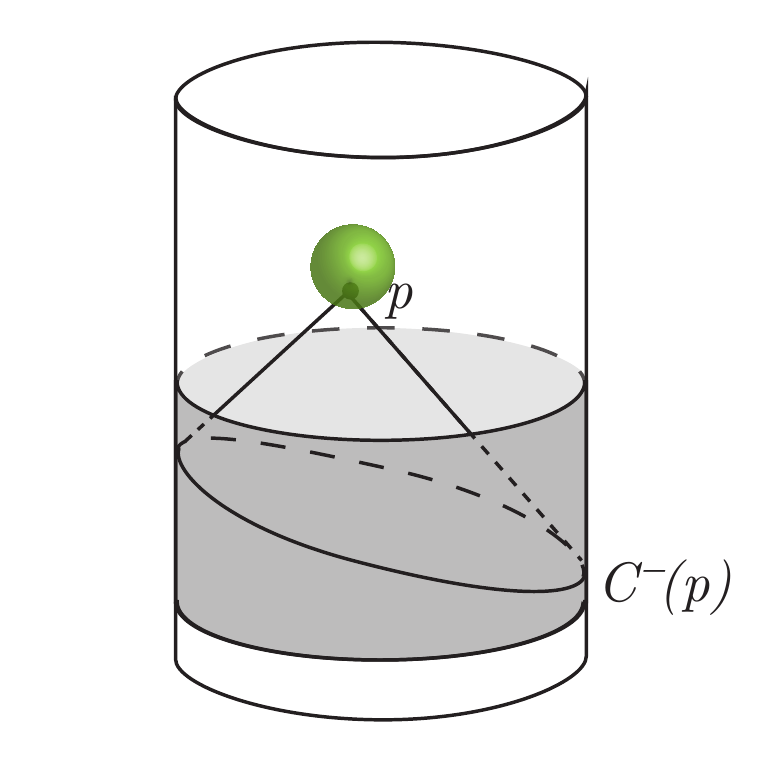}
\caption{The conformal metric of a region deep in the bulk is contained in a time strip of the boundary consisting of cuts of points in the bulk region.}
\label{fig:regionsubregion}
\end{figure} 

Finally, our approach to bulk reconstruction  suggests a new type of subregion/ subregion duality.  This phrase is usually interpreted as meaning that a region of the boundary such as a causal diamond is dual to a bulk region anchored on the asymptotic causal diamond~\cite{BouFre12, BouLei12, CzeKar12, HubRan12}. By considering light-cone cuts, one sees that a time strip of the boundary can describe a region in the domain of influence  deep in the bulk, as illustrated in Fig.~\ref{fig:regionsubregion}. (To obtain the cuts from the correlation functions, one would need two time strips.)  It would be interesting to investigate this different formulation of subregion/subregion duality in more depth.

\end{spacing}
\section*{Acknowlegements}
It is a pleasure to thank S. Fischetti, D. Garfinkle, D. Harlow, S. Hartnoll, J. Maldacena,  D. Marolf, and A. Wall for discussions. 
 This work was supported in part by NSF grant PHY-1504541. The work of NE was also supported by the NSF Graduate Research Fellowship under grant DE-1144085 and by funds from the University of California.

\appendix
\begin{spacing}{1.2}
\section{Theorems and Proofs}\label{appendix}

Here we will state and prove all the results mentioned in this paper. Our assumptions (as stated in Sec.~\ref{sec:goodcut}) are the following: $M$ is at least $C^{2}$, maximally extended, connected, AdS hyperbolic, and asymptotically AdS. Recall that AdS hyperbolic means that there are no closed causal curves, and for any two points $(p,q)$, $J^+(p)\cap J^-(q)$ is compact after conformally compactifying the AdS boundary~\cite{Wal12}. $\partial M$ is maximally extended, connected, and globally hyperbolic. We also assume cuts are $C^1$ everywhere except on a set of measure zero.

An important fact that we will use below is that continuous deformations of bulk points correspond to continuous deformations of their corresponding cuts. This follows from the fact that the cuts are determined by null geodesics which satisfy ODEs, and every null geodesic has an open neighborhood in a maximally extended, globally hyperbolic spacetime.

We will assume everywhere in this section that $p$ and $q$ are bulk points in the domain of influence of the asymptotic boundary, so that $C(p)$ and $C(q)$ are nonempty cuts corresponding to $p$ and $q$. We shall work with past cuts where the generalization to future cuts is obvious.\\

\noindent \textbf{Proposition:} (1) $C(p)$ is a complete spatial slice of the boundary $\partial M$, (2) For every point $p\in I^{+}[\partial M]$, there is precisely one past  cut and for every point $p\in I^{-}[\partial M]$ there is precisely one future cut,  (3)  $C(p)\cap C(q)$ contains a nonempty open set if and only if $ p=q$.\\

\begin{proof}  The proof is in three parts. (1): Since $\partial J^{-}(p)$ is achronal,  any cross-section of it is likewise achronal. The cut $C(p)$ must be spacelike since it is the intersection of $\partial J^{-}(p)$ with a timelike surface.  We prove that a cut $C(p)$ of some bulk point $p$ is a complete slice of $\partial M$ by contradiction: assume that there exists a complete spatial slice  $\Sigma$ of $\partial M$ which contains $C(p)$ as a proper subset. Consider evolving $\Sigma$ slightly backwards in time to a new slice $\Sigma'$ where $\Sigma\cap\Sigma'=\varnothing$. By definition (and by AdS hyperbolicity), $C(p)\subset \partial J^{-}(p)=\partial I^{-}(p)$, so there will be points on $\Sigma'$ that lie in $I^{-}(p)$ (by AdS hyperbolicity, $\partial J^{-}(p)=\partial I^{-}(p)$). Thus every point on $\Sigma'$ which is in the (chronological) past of $C(p)$ lies in $I^{-}(p)$.  By assumption, $C(p)$ is a proper subset of $\Sigma$, so the set $\Sigma - C(p)$ is nonempty. Working in the conformal compactification of $M$, we find by Theorem 8.3.11 of~\cite{Wald} that $J^{-}(p)$ is closed; it follows that $C(p)$, the boundary of $J^{-}(p)$ on $\partial M$, is closed as well. This implies that $\Sigma - C(p)$ cannot consist of isolated points, i.e. there is at least one connected open neighborhood $S \subset \Sigma -C(p)$. By taking $\Sigma'$ very close to $\Sigma$, we can find an open set $S'\subset \Sigma'$ which is everywhere acausal to $p$ (since $S$ is acausal to $p$). Since $\Sigma'$ contains subsets of $I^{-}(p)$ and subsets that are acausal to $p$, it follows that $\Sigma'$ must also contain points in $\partial I^{-}(p)$. But these points are by definition in $C(p)$, in contradiction with the assumption that $C(p)$ lies everywhere on $\Sigma$ and $\Sigma\cap \Sigma'=\varnothing$.\\
(2): If $p\in I^{+}[\partial M]$, there are past-directed causal curves from $p$ to $\partial M$, so $I^{-}(p)\cap \partial M \neq \varnothing$. By AdS hyperbolicity, $\partial M$ cannot lie entirely in the chronological past of a bulk point, so $C^-(p)=\partial J^{-}(p)\cap \partial M\neq \varnothing$. By 
the 
first part of this 
proposition, $C(p)$ is a complete achronal slice of $\partial M$. This proves existence. It is clear that two distinct cuts cannot correspond to the same bulk point (that would require that $\partial J^{-}(p)$ intersect $\partial M$ on two complete achronal slices; because $\partial M$ is timelike, points on those slices would be timelike to one another, in contradiction with the achronality of $\partial J^{-}(p)$). The proof for future cuts is similar. \\
(3): If $p =q$, then clearly $C(p) = C(q)$ so their intersection contains a nonempty open set. We now prove the converse: Let $U= C(p)\cap C(q)$, and let $O\subset U$ be an open $C^{1}$ set (this exists by assumption). Assuming these are past cuts, there is a unique future-directed, inwards-pointing null congruence orthogonal to $O$. Let $\gamma_{1}$ and $\gamma_{2}$ be two generators of this null congruence; so $\gamma_{1}$ and $\gamma_{2}$ are also generators of $\partial J^{-}(p)$ and $\partial J^{-}(q)$. Because $O \subset C(p)$, $\gamma_{1}$ and $\gamma_{2}$ must cross at $p$. Similarly, $\gamma_{1}$ and $\gamma_{2}$ must cross at $q$. But the generators of $\partial J^{-}(p)$ which reach the boundary can have an intersection only at $p$. Similarly, these generators of $\partial J^{-}(q)$ can cross only at $q$. So we find that $p=q$.\\  \end{proof}

\noindent \textbf{Lemma 1:}  If $q\in I^{-}(p)$, then $C^-(q)\subset I^{-}[C^-(p)]$. Similarly, if $p\in I^{+}(q)$, then $C^{+}(p)\subset I^{+}[C^{+}(q)]$.
\begin{proof}
If $q\in I^{-}(p)$, then there exist timelike past-directed paths from $p$ to $q$, and past-directed null paths from $q$ to $C^{-}(q)$. These can be combined to form
 timelike past-directed paths from $p$ to $C^{-}(q)$, so $C^{-}(q)\subset I^{-}(p)$. This immediately implies that $C^{-}(q)\cap C^{-}(p)=\varnothing$. Since any point on $\partial M$ which is to the future of $C^{-}(p)$ is acausal to $p$, we find that $C^{-}(q)$ must lie in the past of $C^{-}(p)$: $C^{-}(q)\subset I^{-}[C^{-}(p)]$. Similarly, we may reverse the argument for future cuts: if  $p\in I^{+}(q)$, and $C^{+}(p)\subset I^{+}[C^{+}(q)]$.

\end{proof}

\noindent \textbf{Lemma 2:} If $q\in \partial J^{-}(p)$, then $C^-(q)\cap C^-(p)$ is at most single point.
\begin{proof}
 Let $x, \ y \in C(q)\cap C(p)$. Then there is a null curve $\gamma_{x}$ from $x$ to $q$ and also a curve $\gamma_{y}$ from $y$ to $q$. Because $q$ is null related to $p$, there is a null curve $\gamma$ from $q$ to $p$. We can therefore join $\gamma_{x}$ to $\gamma$ to obtain a causal curve from $x$ to $p$ and similarly join $\gamma_{y}$ to $\gamma$ to get a causal curve from $y$ to $p$. If either curve is broken, then one of $x$ and $y$ is timelike to $p$, in contradiction with the achronality of $\partial J^{-}(p)$. If neither curve is a broken curve, then $\gamma_{x}=\gamma_{y}$, so $x=y$. 

\end{proof}

\noindent A similar result holds for future cuts.\\

\noindent \textbf{Lemma 3:} If there exists an open subset $O$ of  $C(q)$, where $O\subset J^{-}[C(p)]$ and $S= C(p)\cap O \neq \varnothing$, where both $C(p)$ and $C(q)$ are $C^{1}$ at $S$, then $q$ and $p$ are null-separated.
\begin{proof}
Since (1) $S$ lies in an open set in the past of $C(p)$, and (2) $C(q)$ and $C(p)$ are $C^{1}$ and intersect at $S$, we conclude that $C(q)$ and $C(p)$ are tangent at $S$. $\partial J^{-}(p)$ and $\partial J^{-}(q)$ must therefore share a generator. This can only happen if $q\in \partial J^{-}(p)$ (because $S\subset \partial J^{-}(p)$). 
\end{proof}

We now prove a stronger form of the result we used in section 3.1:\\

\noindent \textbf{Theorem 1:} Let $\gamma$ be an achronal null geodesic from $p$ to a $C^1$ point $r\in C(p)$. Then $q$ is on $\gamma$ if and only if $C(q) \cap C(p) = \{r\}$ and $C(q)$ is also  $C^1$ at $r$.

\begin{proof} If $q$ is on $\gamma$, then $\gamma$ is a generator of $\p J^-(q)$. If there were a second null geodesic $\tilde \gamma$ from $q$ to $r$, one could combine it with $\gamma$ and obtain a broken null geodesic from $p$ to $r$ contradicting the fact that $r\in C^-(p)$.  Hence $r$ is also a regular point of $C(q)$. 
By Lemma 2, this can be the only point of intersection of the cuts. Conversely, if $C(p)$ and $C(q)$ coincide at $r$, then one must lie entirely in the causal past of the other, say $C(q)\subset J^{-}(C(p))$, with a single regular point of intersection. By Lemma 3, $p$ and $q$ are null related, with $q\in \partial J^{-}(p)$. 
\end{proof}

\noindent \textbf{Definition:} $C(p)$ and $C(q)$ {\it cross}  
if $C(p)\cap I^{+}(C(q))\neq \varnothing$ and $C(q)\cap I^{+}(C(p))\neq \varnothing$.\\

All the results above only use one set of cuts, either past or future. If we restrict to points in the causal wedge of the entire boundary, we have  both past and future cuts and can say more:\\

\noindent \textbf{Theorem 2:} $p$ and $q$ are spacelike separated if one of the following is true:
\begin{enumerate}
	\item  $C(p)$ and $C(q)$ cross, where $C(p)$ and $C(q)$ are either both past or both future cuts (Fig.~\ref{subfig:crossing}).
	\item $C^{\pm}(q)$ both lie between $C^{+}(p)$ and $C^{-}(p)$ (Fig.~\ref{subfig:sandwich}).
\end{enumerate}

\begin{proof} (1) If the cuts cross, then by definition, 
$C(p)\cap I^{+}(C(q))\neq \varnothing$ and $C(q)\cap I^{+}(C(p))\neq \varnothing$. The former implies that there are points in $C(p)$ that are outside of causal contact with $q$. Because the generators of $\partial J^{-}(p)$ are causal themselves and $C(p)$ is a complete slice, this implies that there are generators of $\partial J^{-}(p)$ that lie always outside of $J^{-}(q)$. So $p \notin J^{-}(q)$. The same line of reasoning shows that $q\notin J^{-}(p)$.  Therefore, the points are spacelike related.
 (2)   Assume that the cuts of $q$ both lie in between the cuts of $p$. Let $\gamma$ be a smooth curve of constant signature\footnote{By this we mean that the curve is everywhere either spacelike, timelike, or null.}  connecting $q$ and $p$. By deforming $q$ to $p$ along $\gamma$, we must smoothly deform the cuts of $q$ towards the cuts of $p$. If $\gamma$ were causal and future-directed, then by Lemma 1, both cuts of $q$ would move towards the future, which would not result in $C^{-}(q)$ agreeing with $C^{-}(p)$ as $q\rightarrow p$. Similarly, if it was causal and past-directed, both cuts of $q$ would move to the past and  $C^{+}(q)$ would not agree with $C^{+}(p)$. Therefore $\gamma$ must be spacelike. 
\end{proof}

\end{spacing}

\bibliographystyle{JHEP}

\bibliography{all}

\providecommand{\href}[2]{#2}\begingroup\raggedright\begin{thebibliography}{10}

\bibitem{Mal97}
J.~Maldacena, {\it The large {$N$} limit of superconformal field theories and
  supergravity},  {\em Adv. Theor. Math. Phys.} {\bf 2} (1998) 231,
  [\href{http://arxiv.org/abs/hep-th/9711200}{{\tt hep-th/9711200}}].

\bibitem{Wit98a}
E.~Witten, {\it {A}nti-de~{S}itter space and holography},  {\em Adv. Theor.
  Math. Phys.} {\bf 2} (1998) 253,
  [\href{http://arxiv.org/abs/hep-th/9802150}{{\tt hep-th/9802150}}].

\bibitem{GubKle98}
S.~S. Gubser, I.~R. Klebanov, and A.~M. Polyakov, {\it Gauge theory correlators
  from noncritical string theory},  {\em Phys. Lett.} {\bf B428} (1998) 105,
  [\href{http://arxiv.org/abs/hep-th/9802109}{{\tt hep-th/9802109}}].

\bibitem{Van09}
M.~Van~Raamsdonk, {\it {Comments on quantum gravity and entanglement}},
  \href{http://arxiv.org/abs/0907.2939}{{\tt arXiv:0907.2939}}.

\bibitem{Van10}
M.~Van~Raamsdonk, {\it {Building up spacetime with quantum entanglement}},
  {\em Gen. Rel. Grav.} {\bf 42} (2010) 2323--2329,
  [\href{http://arxiv.org/abs/1005.3035}{{\tt arXiv:1005.3035}}]. [Int. J. Mod.
  Phys.D19,2429(2010)].

\bibitem{RyuTak06}
S.~Ryu and T.~Takayanagi, {\it {Holographic derivation of entanglement entropy
  from AdS/CFT}},  {\em Phys.Rev.Lett.} {\bf 96} (2006) 181602,
  [\href{http://arxiv.org/abs/hep-th/0603001}{{\tt hep-th/0603001}}].

\bibitem{HubRan07}
V.~E. Hubeny, M.~Rangamani, and T.~Takayanagi, {\it {A Covariant holographic
  entanglement entropy proposal}},  {\em JHEP} {\bf 0707} (2007) 062,
  [\href{http://arxiv.org/abs/0705.0016}{{\tt arXiv:0705.0016}}].

\bibitem{BalChoCze}
V.~Balasubramanian, B.~D. Chowdhury, B.~Czech, J.~de~Boer, and M.~P. Heller,
  {\it {Bulk curves from boundary data in holography}},  {\em Phys.Rev.} {\bf
  D89} (2014), no.~8 086004, [\href{http://arxiv.org/abs/1310.4204}{{\tt
  arXiv:1310.4204}}].

\bibitem{MyeHea14}
M.~Headrick, R.~C. Myers, and J.~Wien, {\it {Holographic Holes and Differential
  Entropy}},  {\em JHEP} {\bf 1410} (2014) 149,
  [\href{http://arxiv.org/abs/1408.4770}{{\tt arXiv:1408.4770}}].

\bibitem{CzeLam}
B.~Czech and L.~Lamprou, {\it {Holographic definition of points and
  distances}},  {\em Phys.Rev.} {\bf D90} (2014), no.~10 106005,
  [\href{http://arxiv.org/abs/1409.4473}{{\tt arXiv:1409.4473}}].

\bibitem{CzeLamMcC15a}
B.~Czech, L.~Lamprou, S.~McCandlish, and J.~Sully, {\it {Integral Geometry and
  Holography}},  {\em JHEP} {\bf 10} (2015) 175,
  [\href{http://arxiv.org/abs/1505.05515}{{\tt arXiv:1505.05515}}].

\bibitem{BalCzeCho}
V.~Balasubramanian, B.~Czech, B.~D. Chowdhury, and J.~de~Boer, {\it {The
  entropy of a hole in spacetime}},  {\em JHEP} {\bf 1310} (2013) 220,
  [\href{http://arxiv.org/abs/1305.0856}{{\tt arXiv:1305.0856}}].

\bibitem{CzeDonSul}
B.~Czech, X.~Dong, and J.~Sully, {\it {Holographic Reconstruction of General
  Bulk Surfaces}},  {\em JHEP} {\bf 1411} (2014) 015,
  [\href{http://arxiv.org/abs/1406.4889}{{\tt arXiv:1406.4889}}].

\bibitem{MyeRaoSug}
R.~C. Myers, J.~Rao, and S.~Sugishita, {\it {Holographic Holes in Higher
  Dimensions}},  {\em JHEP} {\bf 1406} (2014) 044,
  [\href{http://arxiv.org/abs/1403.3416}{{\tt arXiv:1403.3416}}].

\bibitem{Hub12}
V.~E. Hubeny, {\it {Extremal surfaces as bulk probes in AdS/CFT}},  {\em JHEP}
  {\bf 1207} (2012) 093, [\href{http://arxiv.org/abs/1203.1044}{{\tt
  arXiv:1203.1044}}].

\bibitem{EngWal13}
N.~Engelhardt and A.~C. Wall, {\it {Extremal Surface Barriers}},  {\em JHEP}
  {\bf 1403} (2014) 068, [\href{http://arxiv.org/abs/1312.3699}{{\tt
  arXiv:1312.3699}}].

\bibitem{EngFis15}
N.~Engelhardt and S.~Fischetti, {\it {Covariant Constraints on Hole-ography}},
  {\em Class. Quant. Grav.} {\bf 32} (2015), no.~19 195021,
  [\href{http://arxiv.org/abs/1507.00354}{{\tt arXiv:1507.00354}}].

\bibitem{deHSolSke00}
S.~de~Haro, S.~N. Solodukhin, and K.~Skenderis, {\it {Holographic
  reconstruction of space-time and renormalization in the AdS / CFT
  correspondence}},  {\em Commun. Math. Phys.} {\bf 217} (2001) 595--622,
  [\href{http://arxiv.org/abs/hep-th/0002230}{{\tt hep-th/0002230}}].

\bibitem{HamKab06}
A.~Hamilton, D.~N. Kabat, G.~Lifschytz, and D.~A. Lowe, {\it {Holographic
  representation of local bulk operators}},  {\em Phys.Rev.} {\bf D74} (2006)
  066009, [\href{http://arxiv.org/abs/hep-th/0606141}{{\tt hep-th/0606141}}].

\bibitem{Kab11}
D.~Kabat, G.~Lifschytz, and D.~A. Lowe, {\it {Constructing local bulk
  observables in interacting AdS/CFT}},  {\em Phys.Rev.} {\bf D83} (2011)
  106009, [\href{http://arxiv.org/abs/1102.2910}{{\tt arXiv:1102.2910}}].

\bibitem{ChrSke16}
A.~Christodoulou and K.~Skenderis, {\it {Holographic Construction of Excited
  CFT States}},  {\em JHEP} {\bf 04} (2016) 096,
  [\href{http://arxiv.org/abs/1602.02039}{{\tt arXiv:1602.02039}}].

\bibitem{MalSimZhi}
J.~Maldacena, D.~Simmons-Duffin, and A.~Zhiboedov, {\it {Looking for a bulk
  point}},  \href{http://arxiv.org/abs/1509.03612}{{\tt arXiv:1509.03612}}.

\bibitem{New76}
E.~T. Newman, {\it {Heaven and Its Properties}},  {\em Gen. Rel. Grav.} {\bf 7}
  (1976) 107--111.

\bibitem{HanNewPen}
R.~O. Hansen, E.~T. Newman, R.~Penrose, and K.~P. Tod, {\it {The Metric and
  Curvature Properties of H Space}},  {\em Proc. Roy. Soc. Lond.} {\bf A363}
  (1978) 445--468.

\bibitem{KozNew83}
C.~N. Kozameh and E.~T. Newman, {\it {Theory of light cone cuts of null
  infinity}},  {\em J. Math. Phys.} {\bf 24} (1983) 2481--2489.

\bibitem{PolSus99}
J.~Polchinski, L.~Susskind, and N.~Toumbas, {\it {Negative energy,
  superluminosity and holography}},  {\em Phys.Rev.} {\bf D60} (1999) 084006,
  [\href{http://arxiv.org/abs/hep-th/9903228}{{\tt hep-th/9903228}}]. Expanded
  version replacing earlier hep-th 9902182.

\bibitem{GarGid09}
M.~Gary, S.~B. Giddings, and J.~Penedones, {\it {Local bulk S-matrix elements
  and CFT singularities}},  {\em Phys. Rev.} {\bf D80} (2009) 085005,
  [\href{http://arxiv.org/abs/0903.4437}{{\tt arXiv:0903.4437}}].

\bibitem{HeePen09}
I.~Heemskerk, J.~Penedones, J.~Polchinski, and J.~Sully, {\it {Holography from
  Conformal Field Theory}},  {\em JHEP} {\bf 10} (2009) 079,
  [\href{http://arxiv.org/abs/0907.0151}{{\tt arXiv:0907.0151}}].

\bibitem{Pen10}
J.~Penedones, {\it {Writing CFT correlation functions as AdS scattering
  amplitudes}},  {\em JHEP} {\bf 03} (2011) 025,
  [\href{http://arxiv.org/abs/1011.1485}{{\tt arXiv:1011.1485}}].

\bibitem{OkuPen11}
T.~Okuda and J.~Penedones, {\it {String scattering in flat space and a scaling
  limit of Yang-Mills correlators}},  {\em Phys. Rev.} {\bf D83} (2011) 086001,
  [\href{http://arxiv.org/abs/1002.2641}{{\tt arXiv:1002.2641}}].

\bibitem{Wal12}
A.~C. Wall, {\it {Maximin Surfaces, and the Strong Subadditivity of the
  Covariant Holographic Entanglement Entropy}},  {\em Class.Quant.Grav.} {\bf
  31} (2014), no.~22 225007, [\href{http://arxiv.org/abs/1211.3494}{{\tt
  arXiv:1211.3494}}].

\bibitem{Wald}
R.~M. Wald, {\em General Relativity}.
\newblock The University of Chicago Press, Chicago, 1984.

\bibitem{HSpaceRev}
M.~Ko, M.~Ludvigsen, E.~T. Newman, and K.~P. Tod, {\it {The theory of H
  -space}},  {\em Phys.Rept.} {\bf 71, 51} (1981).

\bibitem{MalPC}
J.~Maldacena, private communication.

\bibitem{GaoWal00}
S.~Gao and R.~M. Wald, {\it {Theorems on gravitational time delay and related
  issues}},  {\em Class. Quant. Grav.} {\bf 17} (2000) 4999--5008,
  [\href{http://arxiv.org/abs/gr-qc/0007021}{{\tt gr-qc/0007021}}].

\bibitem{FreHub05}
B.~Freivogel, V.~E. Hubeny, A.~Maloney, R.~C. Myers, M.~Rangamani, and
  S.~Shenker, {\it {Inflation in AdS/CFT}},  {\em JHEP} {\bf 03} (2006) 007,
  [\href{http://arxiv.org/abs/hep-th/0510046}{{\tt hep-th/0510046}}].

\bibitem{SheSta14}
S.~H. Shenker and D.~Stanford, {\it {Black holes and the butterfly effect}},
  {\em JHEP} {\bf 03} (2014) 067, [\href{http://arxiv.org/abs/1306.0622}{{\tt
  arXiv:1306.0622}}].

\bibitem{CzeKar12}
B.~Czech, J.~L. Karczmarek, F.~Nogueira, and M.~Van~Raamsdonk, {\it {The
  Gravity Dual of a Density Matrix}},  {\em Class.Quant.Grav.} {\bf 29} (2012)
  155009, [\href{http://arxiv.org/abs/1204.1330}{{\tt arXiv:1204.1330}}].

\bibitem{HeaHub14}
M.~Headrick, V.~E. Hubeny, A.~Lawrence, and M.~Rangamani, {\it {Causality \&
  holographic entanglement entropy}},  {\em JHEP} {\bf 12} (2014) 162,
  [\href{http://arxiv.org/abs/1408.6300}{{\tt arXiv:1408.6300}}].

\bibitem{JafSuh14}
D.~L. Jafferis and S.~J. Suh, {\it {The Gravity Duals of Modular
  Hamiltonians}},  \href{http://arxiv.org/abs/1412.8465}{{\tt
  arXiv:1412.8465}}.

\bibitem{JafLew15}
D.~L. Jafferis, A.~Lewkowycz, J.~Maldacena, and S.~J. Suh, {\it {Relative
  entropy equals bulk relative entropy}},
  \href{http://arxiv.org/abs/1512.06431}{{\tt arXiv:1512.06431}}.

\bibitem{DonHar16}
X.~Dong, D.~Harlow, and A.~C. Wall, {\it {Bulk Reconstruction in the
  Entanglement Wedge in AdS/CFT}},  \href{http://arxiv.org/abs/1601.05416}{{\tt
  arXiv:1601.05416}}.

\bibitem{BouFre12}
R.~Bousso, B.~Freivogel, S.~Leichenauer, V.~Rosenhaus, and C.~Zukowski, {\it
  {Null Geodesics, Local CFT Operators and AdS/CFT for Subregions}},  {\em
  Phys.Rev.} {\bf D88} (2013) 064057,
  [\href{http://arxiv.org/abs/1209.4641}{{\tt arXiv:1209.4641}}].

\bibitem{BouLei12}
R.~Bousso, S.~Leichenauer, and V.~Rosenhaus, {\it {Light-sheets and AdS/CFT}},
  {\em Phys.Rev.} {\bf D86} (2012) 046009,
  [\href{http://arxiv.org/abs/1203.6619}{{\tt arXiv:1203.6619}}].

\bibitem{HubRan12}
V.~E. Hubeny and M.~Rangamani, {\it {Causal Holographic Information}},  {\em
  JHEP} {\bf 1206} (2012) 114, [\href{http://arxiv.org/abs/1204.1698}{{\tt
  arXiv:1204.1698}}].

\end{thebibliography}\endgroup

\end{document}